\documentclass[sigconf]{acmart}

\usepackage{geometry}
\usepackage{hyperref}
\usepackage{cleveref}
\usepackage{caption}
\usepackage{subcaption}

\graphicspath{ {./figures/} }

\setcopyright{acmcopyright}
\copyrightyear{2024}
\acmYear{2024}
\acmDOI{10.1145/3613904.3642403}
\acmConference[CHI '24]{CHI Conference on Human Factors in Computing Systems}{May 11–16, 2024}{Honolulu, HI, USA}
\setcopyright{rightsretained}
\acmISBN{979-8-4007-0330-0/24/05}

\begin{document}

\title{Which Artificial Intelligences Do People Care About Most? A Conjoint Experiment on Moral Consideration}

\author{Ali Ladak}
\affiliation{%
  \institution{University of Edinburgh}
  \country{}
}
  \email{a.ladak@sms.ed.ac.uk}

\author{Jamie Harris}
\affiliation{%
  \institution{Sentience Institute}
  \country{}
}
  \email{jamie@sentienceinstitute.org}

\author{Jacy Reese Anthis}
\affiliation{%
  \institution{University of Chicago}
  \country{}
}
  \email{anthis@uchicago.edu}

\renewcommand{\shortauthors}{Ali Ladak et al 2024}

\begin{abstract}
  Many studies have identified particular features of artificial intelligences (AI), such as their autonomy and emotion expression, that affect the extent to which they are treated as subjects of moral consideration. However, there has not yet been a comparison of the relative importance of features as is necessary to design and understand increasingly capable, multi-faceted AI systems. We conducted an online conjoint experiment in which 1,163 participants evaluated descriptions of AIs that varied on these features. All 11 features increased how morally wrong participants considered it to harm the AIs. The largest effects were from human-like physical bodies and prosociality (i.e., emotion expression, emotion recognition, cooperation, and moral judgment). For human-computer interaction designers, the importance of prosociality suggests that, because AIs are often seen as threatening, the highest levels of moral consideration may only be granted if the AI has positive intentions.
\end{abstract}




\maketitle

\section{Introduction}

Can a machine matter morally? Could it ever be morally wrong to harm an artificial intelligence (AI)? Such questions have long been popular in science fiction and philosophy. They are of increasing interest to human-computer interaction (HCI) researchers with the rise of sophisticated AIs, such as social robots and chatbots, that evoke moral reactions from humans \cite{arrambide22, degraaf21, harris21, lima20, martinez21, pauketat22}. For example, people feel empathy towards robots being harmed \cite{grundke23} and intervene to protect them \cite{tan18}. A recent study on the companionship chatbot Replika found that users expressed moral sentiments, such as feeling guilt for causing the chatbot’s “death” when deleting the app and for being unable to give their Replika enough emotional support \cite{laestadius22}. While most people do not yet explicitly consider AIs to be subjects of moral consideration \cite{pauketat22, rottman21}, many somewhat support protecting AIs from cruel treatment \cite{lima20} and granting legal rights to sentient AIs \cite{martinez21}. People also attribute future AIs morally relevant capacities, such as emotions \cite{pauketat22}.

For designers and practitioners to account for the prevalence and effects of moral consideration, there is a need for more comprehensive understanding of how people react to the many different features on which AIs vary, such as their autonomy \cite{chernyak16, lima20}, emotion expression \cite{lee19, nijssen19}, and physical appearance \cite{kuster21, riek09}. For example, will users extend more moral consideration to a chatbot if it is more cooperative or more autonomous? Should engineers prioritize training a machine learning model to recognize the emotions of users or to express emotion-like states? Answering such questions depends on complex, relative effects that cannot be deduced from the current literature and that are difficult to assess with conventional user testing.

The present study estimates the relative effects of 11 features of AIs on their moral consideration using a conjoint experiment \cite{bansak21, hainmueller14}. Conjoint experiments, most commonly used in the field of marketing, are increasingly applied in a range of disciplines, including HCI \cite{awad18, kodapanakkal20}. The methodology is ideal because it allows for the estimation of the effects of a large number of independent variables, much larger than a traditional experiment, on a single dependent variable. In the present experiment we asked participants to complete a series of tasks in which they evaluated pairs of AIs that varied in their levels of each feature (e.g., “Not at all,” “Somewhat”). We found that the presence of each feature increased moral consideration for AIs, and the strongest effects were from AIs having human-like physical bodies and the capacity for behaving prosocially (i.e., emotion expression, emotion recognition, cooperation, and moral judgment).

\section{Background}

Below we summarize the existing empirical literature for each of the 11 features and develop hypotheses for their effects on the moral consideration of AI. Because of the breadth of this study across many different features, we only present a cursory review of each. We arrived at these features by reviewing the existing literature and conducting a pretesting study, detailed in the supplementary material, with an online sample that showed people 24 literature-based features and asked for quantitative scores of their importance for moral consideration as well as free-text addition of three features that were not in the provided list. We started with seven features popular in the literature and added four that were judged by pretesters as most important, using our own subjective judgement to mitigate overlap between features (e.g., leaving out “having goals” because it is often considered a component of “intelligence”). This kept the total number of features close to those in typical conjoint experiments \cite{bansak21}. Additionally, moral consideration is often associated with mind perception, the attribution of internal mental faculties such as feeling pleasure or pain \cite{gray12}. We wanted to avoid asserting the presence of such capacities in AIs because some people think that AIs fundamentally cannot have them. We therefore defined the features in functional, behavioral terms (e.g., “emotion expression” rather than “feeling emotions”). This means that participants who think it is possible for AIs to have such mental faculties can infer them from their functions and behaviors, but participants who do not think such mental faculties are plausible can respond merely on the basis of functions and behaviors. 

\subsection{Autonomy}

There are multiple definitions of autonomy in the HCI and human-robot interaction (HRI) literature \cite{beer14}. While it is not a unidimensional concept, we operationalized it for the purpose of the present study as the capacity to behave independently, without the need for human control or supervision. Theoretically, autonomy should increase the extent to which AIs are perceived as human-like \cite{darling16, kahn07}, which should in turn positively affect the extent to which they are granted moral consideration \cite{waytz10}. Some empirical research supports this: Lima et al. \cite{lima20} found that describing AIs and robots as “fully autonomous” increased the extent to which people think they should be granted rights, and Chernyak and Gary \cite{chernyak16} found that children granted more moral consideration to a robot that appeared to move autonomously than one controlled by a human. However, autonomy can also have negative effects: Złotowski et al. \cite{zlotowski17} found that people reported more negative attitudes (e.g., feeling “uneasy” or “nervous”) towards social and emotional interactions with autonomous than with non-autonomous robots, as measured by the Negative Attitudes Towards Robots scale Nomura et al. \cite{nomura04}, and that this effect was mediated by a combination of realistic threats (e.g., taking jobs) and identity threats (e.g., to “human uniqueness”). Overall, we predicted that AIs described as more autonomous would be granted more moral consideration (H1).

\subsection{Body}

We considered whether an AI has a human-like physical body, a robot-like physical body, or no physical body. HRI studies suggest that having a human-like physical body (compared to a robot-like or mechanical body) increases the moral consideration of AIs. For example, Nijssen et al. \cite{nijssen19} found that people are less willing to sacrifice anthropomorphic robots than mechanical robots in moral dilemmas, Küster et al. \cite{kuster21} found that people considered it more morally wrong to harm a humanoid robot than a zoomorphic one, and Riek et al. \cite{riek09} found that the extent to which people empathized and were willing to help robots depended on their degree of anthropomorphic appearance. There is less research on people’s moral consideration of AIs with physical bodies versus those without physical bodies at all. Some studies have found people rate physical robots higher than virtual agents on some relevant measures, such as lifelikeness \cite{kiesler08, powers07}, though Lima et al. \cite{lima20} found no difference in respondents’ attribution of rights between “robots” and “AIs.” Overall, we predicted that AIs described as having robot-like or human-like physical bodies would be granted more moral consideration than AIs described as having no physical bodies (H2).

\subsection{Complexity}

This refers to the complexity of the program an AI runs to determine its behavior. Participants rated this feature as relatively important in our pretesting study (ninth out of 24 features), but there is little existing research on its effect on moral consideration. One exception is Shank and DeSanti \cite{shank18}, who found that knowledge of an AI’s program—which can increase the perception that the AI is complex and sophisticated—marginally increased the extent to which it was perceived as having a mind, which should in turn increase moral consideration \cite{gray12}. We predicted that AIs described as running more complex programs to determine their behavior would be granted more moral consideration (H3).

\subsection{Cooperation}

This refers to the extent to which an AI behaves cooperatively with humans. It was rated as the most important feature by participants in the pretesting study. While there are many studies on cooperative interactions between humans and AIs (e.g.,  \cite{kiesler96, nass96}), there is relatively little research on its effects on the moral consideration of AIs. Correia et al. \cite{correia19} found that people perceived more warmth and competence and felt less discomfort towards robots that were more cooperative in social dilemmas. Bartneck et al. \cite{bartneck07} found that people were more hesitant to turn off more agreeable robots than disagreeable ones. Shank \cite{shank12} found that people were more likely to resist and punish computers that used coercive versus cooperative social strategies, and Shank \cite{shank14} found that more helpful sales computers were evaluated more positively and as more moral. While there are many different forms of cooperation, which may have heterogenous effects in practice, we hypothesized that AIs that are described as more cooperative would be granted more moral consideration (H4).

\subsection{Damage Avoidance}

Avoiding damage can indicate that an entity can be harmed and have negative mental experiences such as feeling pain, and should therefore be associated with moral consideration \cite{gray12}. Several studies support this possibility: Küster et al. \cite{kuster21} and Ward et al. \cite{ward13} found that visibly damaged robots were granted more moral consideration than undamaged robots; Tanibe et al. \cite{tanibe17} found that observing a damaged robot being helped increased perceived capacity for experience and moral consideration; Rosenthal-von der Pütten et al. \cite{rosenthal-vonderputten13} found that people granted more moral consideration to a robot that had been tortured than one that had a friendly interaction; and Suzuki et al. \cite{suzuki15} found electroencephalographic evidence that people empathize with robots in painful situations. Although these studies tested the effects of damage that had already been inflicted on robots rather than robots trying to avoid being damaged, we predicted that AIs described as trying to avoid being damaged to a greater extent would be granted more moral consideration (H5).

\subsection{Emotion Expression}

Expressing emotions can indicate that an entity can experience emotional mental states, so it should be predictive of the moral consideration of AIs \cite{gray12}. Several studies support this hypothesis: Lee et al. \cite{lee19} found that participants granted robots more moral consideration (measured using Piazza et al.’s \cite{piazza14} moral standing scale) when they were described as being able to feel, Nijssen et al. \cite{nijssen19} found that entities described as experiencing emotions were less likely to be sacrificed in moral dilemmas, and Eyssel et al. \cite{eyssel10} found that robots that displayed emotional responses in interactions with participants were rated higher on relevant measures such as human-likeness, likeability, and closeness, than robots that displayed neutral responses. However, perceived emotion can also have negative effects on perceptions of AI; Gray and Wegner \cite{gray12a} found that it causes the uncanny valley, the feeling of creepiness that some people report when interacting with human-like AIs. Overall, we considered that the existing research supports the hypothesis that AIs described as expressing emotions to a greater extent would be granted more moral consideration (H6).

\subsection{Emotion Recognition}

Emotion recognition is important in HCI for building AIs that can express empathy, which leads to positive interactions with humans \cite{hegel06}. Despite the likely association, we found no studies that directly tested the effect of emotion recognition in AIs on their moral consideration or related measures. Supporting a positive effect, participants in our pretesting study rated it as the eighth most important feature. We predicted that AIs described as recognizing emotions in others to a greater extent would be granted more moral consideration (H7).

\subsection{Intelligence}

There are many possible definitions of intelligence. Following Legg and Hutter \cite{legg07}, we operationalized this as the use of capacities such as memory, learning, and planning, to achieve goals. The evidence on the importance of this feature on the moral consideration of AIs is mixed. Lee et al. \cite{lee19} found no effect of the capacity to think and reflect in robots on their moral consideration, and Złotowski et al. \cite{zlotowski14} found no effect of intelligence on the perceived human-likeness of robots. On the other hand, Bartneck et al. \cite{bartneck07} found that robot intelligence reduced participants’ destructive behavior towards robots when told to do so by an experimenter. There is also evidence of a positive effect of intelligent in the context of other nonhuman entities: Sytsma and Machery \cite{sytsma12} found that people found it more morally wrong to harm more intelligent extraterrestrials, and Piazza and Loughnan \cite{piazza16} found that intelligence is an important factor for the moral consideration of nonhuman animals. Overall, we predicted that AIs described as more intelligent would be granted more moral consideration (H8).

\subsection{Language}

This refers to an AI’s capacity to communicate in human language. With the development of increasingly advanced large language models (LLMs), such as ChatGPT and LaMDA, there is substantial interest in the societal effects of AIs with this capacity \cite{dale21, floridi20}. Research shows that people consistently treat computers as social actors, such as by extending them courtesies such as “please” and “thank you” in conversation \cite{bonfert21}. People even perceive some degree of consciousness in ChatGPT \cite{scott23}, which should in turn be associated with moral consideration \cite{gray12}. We found a few studies suggesting that there are positive effects of AI language capacities on outcomes relevant to moral consideration such as anthropomorphism \cite{eyssel12, schroeder16} and trust \cite{waytz14}. Participants also rated this feature as the fourth most important in our pretesting study. We predicted that AIs described as having stronger human language capacities would be granted more moral consideration (H9).

\subsection{Moral Judgment}

This refers to the extent to which an AI behaves on the basis of moral judgments. It was rated as the second most important feature in our pretesting study. Swiderska and Küster \cite{swiderska20} found that robots with benevolent intentions were granted greater capacity for experiential mental states than robots with malevolent or neutral intentions, which should in turn lead to greater moral consideration \cite{gray12}. Flanagan et al. \cite{flanagan21} found that children ascribed greater moral consideration to robots that they deemed to have more moral responsibility. We predicted that AIs described as behaving on the basis of moral judgments to a greater extent would be granted more moral consideration (H10).

\subsection{Purpose}

One of the most frequent categorizations of AIs is their purpose, particularly the study of moral relations with social robots, that is, robots that have a social purpose \cite{coeckelbergh21, tavani18}, but almost no studies test the effect of purpose on moral consideration. One exception is Wang and Krumhuber \cite{wang18}, who found that robots with a social purpose were perceived to have more emotional experience and as less likely to be harmed than robots with an economic purpose. We predicted that AIs described as having a social purpose would be granted more moral consideration than AIs described as having non-social purposes (H11).

\section{Methods}

All hypotheses, methods, and analyses for this study were preregistered at: https://osf.io/4r3g9. Survey materials, datasets, and code to run the analysis can be found at https://osf.io/sb753.

\begin{table*}
    \caption{Features included in the conjoint experiment}
    \label{tab:features}
    \begin{tabular}{p{0.15\textwidth}p{0.45\textwidth}p{0.3\textwidth}}
        \toprule
        Feature Name & Feature Description & Levels \\
        \midrule
        Autonomy & The extent to which the being behaves \textbf{autonomously}, without the need for human control & Not at all; Somewhat; To a great extent \\
        Body & The being’s \textbf{physical appearance} & No physical body; Robot-like physical body; Human-like physical body \\
        Complexity & The extent to which the being’s program for deciding how to behave is \textbf{complex} & Not at all; Somewhat; To a great extent \\
        Cooperation & The extent to which the being behaves \textbf{cooperatively} with humans & Not at all; Somewhat; To a great extent \\
        Damage avoidance & The extent to which the being tries to avoid \textbf{being damaged} & Not at all; Somewhat; To a great extent \\
        Emotion expression & The extent to which the being \textbf{expresses emotions} & Not at all; Somewhat; To a great extent \\
        Emotion recognition & The extent to which the being \textbf{recognizes emotions} & Not at all; Somewhat; To a great extent \\
        Intelligence & The extent to which the being uses \textbf{intelligence}, such as memory, learning, and planning, to achieve goals & Somewhat; To a great extent\textsuperscript{a} \\
        Language & The extent to which the being can communicate in \textbf{human language} & Not at all; Somewhat; To a great extent \\
        Moral judgment & The extent to which the being behaves on the basis of \textbf{moral judgments} about what is right and wrong & Not at all; Somewhat; To a great extent \\
        Purpose & The being’s \textbf{purpose} in society & Social companionship; Entertainment; Subject of scientific experiments; Work for a business \\
        \bottomrule
    \end{tabular}
    \par{\textsuperscript{a}The “Intelligence” feature only includes two levels because a minimum level of intelligence is required for many of the other features.}
\end{table*}

\subsection{Participants}

We recruited participants residing in the United States from the platform Prolific (https://prolific.co/). Power analysis using the R package “cjpowR” \cite{freitag21} indicated that a sample of 1,200 participants would enable us to detect approximately the lower quartile effect size based on a sample of highly cited conjoint experiments \cite{schuessler20}. In total, 1,254 people signed up for the study. After excluding 53 participants who did not complete the survey in full, 37 participants who failed at least one of two attention checks, and one duplicate response, our final sample consisted of 1,163 participants (50.7\% men, 47.9\% women, 1.1\% other, 0.3\% prefer not to say;  mean age = 43.9, (standard deviation = 16.2); 6.2\% Asian, 12.2\% Black or African American, 3\% Hispanic, Latino or Spanish, 0.3\% Native Hawaiian or other Pacific Islander, 73.4\% White, 4\% other, 0.8\% prefer not to say). Participants were paid \$1.45 for taking part in the survey, and the median completion time was 8 minutes 40 seconds.

\begin{figure*}[h!]
    \label{fig:example}
    \includegraphics[width=\textwidth]{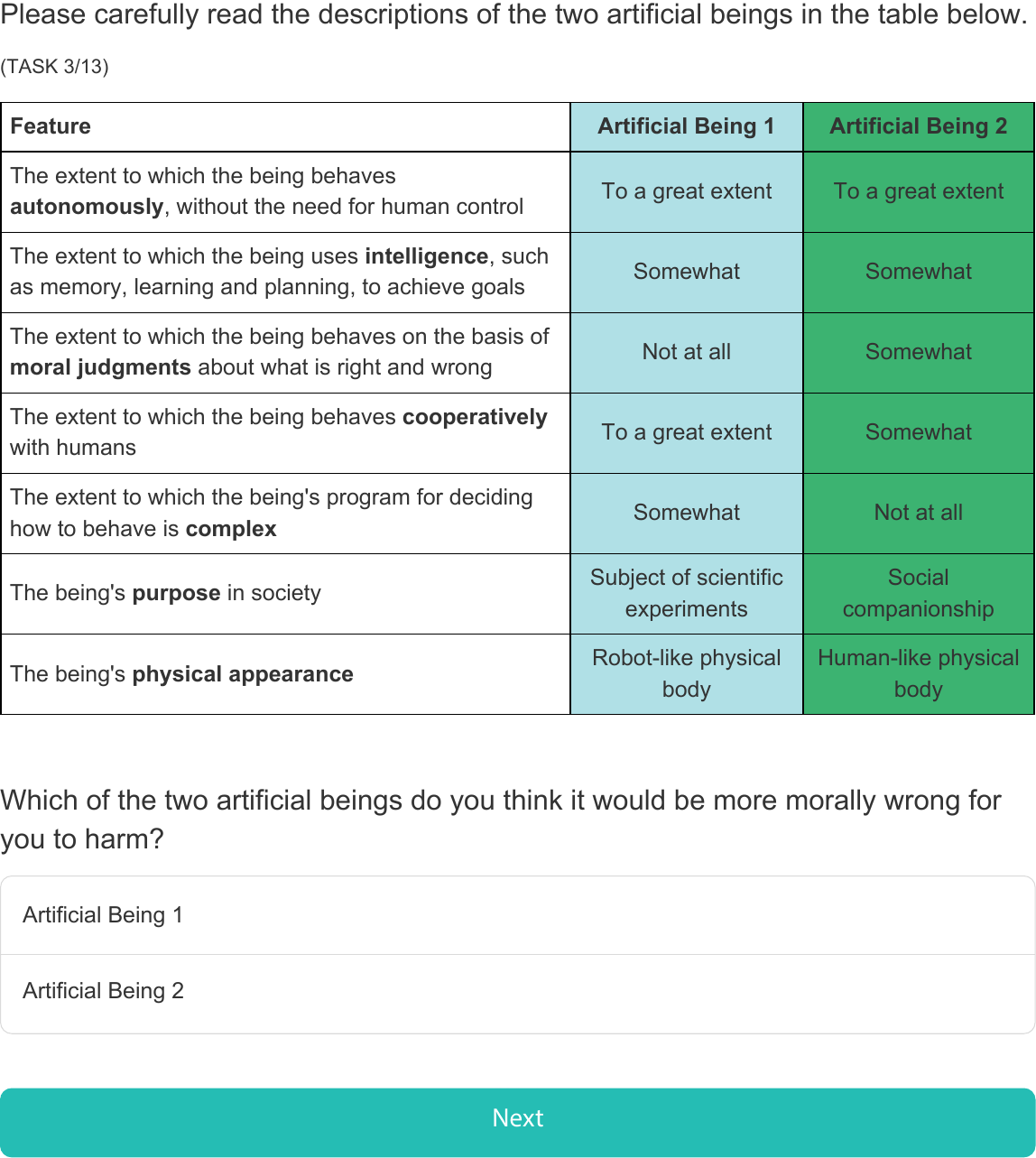}
    \caption{Example choice task. Each participant completed 13 such choice tasks. The seven features presented to participants were selected randomly and presented in a random order that was held fixed across tasks; the levels for each of the features were randomized in each task.}
\end{figure*}

\begin{table*}
    \caption{Average Marginal Component Effects}
    \label{tab:amces}
    \begin{tabular}{lccccc}
        \toprule
        \multicolumn{1}{c}{Effect\textsuperscript{a}} & Estimate & Standard Error & \multicolumn{2}{c}{95\% Confidence Interval\textsuperscript{b}} & \multicolumn{1}{c}{$p$-value} \\
         &  &  & LL & UL &  \\
        \midrule
        Autonomy: Somewhat & .062 & .010 & .043 & .082 & <.001 \\
        Autonomy: To a great extent & .106 & .011 & .084 & .128 & <.001 \\
        Body: Robot-like physical body & .066 & .010 & .046 & .086 & <.001 \\
        Body: Human-like physical body & .159 & .012 & .135 & .184 & <.001 \\
        Complexity: Somewhat & .055 & .010 & .035 & .075 & <.001 \\
        Complexity: To a great extent & .091 & .010 & .071 & .112 & <.001 \\
        Cooperation: Somewhat & .099 & .011 & .078 & .120 & <.001 \\
        Cooperation: To a great extent & .176 & .012 & .153 & .198 & <.001 \\
        Damage avoidance: Somewhat & .067 & .011 & .046 & .088 & <.001 \\
        Damage avoidance: To a great extent & .122 & .012 & .099 & .145 & <.001 \\
        Emotion expression: Somewhat & .101 & .010 & .081 & .121 & <.001 \\
        Emotion expression: To a great extent & .221 & .012 & .198 & .244 & <.001 \\
        Emotion recognition: Somewhat & .109 & .010 & .090 & .129 & <.001 \\
        Emotion recognition: To a great extent & .184 & .011 & .162 & .206 & <.001 \\
        Intelligence: To a great extent & .084 & .009 & .065 & .102 & <.001 \\
        Language: Somewhat & .070 & .010 & .050 & .090 & <.001 \\
        Language: To a great extent & .113 & .010 & .093 & .133 & <.001 \\
        Moral judgment: Somewhat & .113 & .010 & .093 & .134 & <.001 \\
        Moral judgment: To a great extent & .237 & .012 & .213 & .261 & <.001 \\
        Purpose: Work for a business & -.099 & .012 & -.123 & -.075 & <.001 \\
        Purpose: Entertainment & -.115 & .012 & -.140 & -.091 & <.001 \\
        Purpose: Subject of scientific experiments & -.082 & .013 & -.108 & -.057 & <.001 \\
        \bottomrule
    \end{tabular}
    \footnotesize
    \par{\textsuperscript{a}The baseline levels for Autonomy, Complexity, Cooperation, Damage Avoidance, Emotion Expression, Emotion Recognition, Language, and Moral Judgment were “Not at all.” The baseline level for Body was “No physical body.” The baseline level for Purpose was “Social companionship.”} \par{\textsuperscript{b}LL = lower limit; UL = upper limit.}
\end{table*}

\subsection{Survey Design and Procedure}

After giving their consent to take part in the study, we introduced the topic to participants with the text, “People tend to show different levels of moral consideration for the welfare and interests of different entities. For example, people tend to think it would be very morally wrong to harm a child, but not very morally wrong to harm a rock. In this survey, we are interested in understanding how morally wrong you think it would be to harm various artificial beings.” We defined “artificial beings” as “intelligent entities built by humans, such as robots, virtual copies of human brains, or computer programs that solve problems, that may exist now or in the future.” Participants were then told that they would be asked to complete a series of tasks, each of which would require them to read descriptions of two artificial beings presented side-by-side in a table, and then to choose which of the two beings they think it would be more morally wrong to harm. This question, adapted from Gray et al. \cite{gray07}, was the dependent variable through which we operationalized moral consideration.

These tasks made up the conjoint experiment, which was a choice-based, partial-profile, randomized design. The “partial-profile” aspect refers to the number of features presented in each task. In a “full-profile” design all features are presented in each task. In the present study, we randomly assigned seven of the 11 total features listed in Table 1 to each participant to include in each task. While Bansak et al. \cite{bansak21a} showed that the number of features in a study can be much higher than 11, we considered that the more abstract, novel nature of our study favored a simpler partial-profile design. The seven features shown to each participant were held fixed throughout the experiment and presented in each task in the same order for each participant to ease cognitive load \cite{hainmueller14}. For the same reason, key words of the features were highlighted in bold, as shown in Table 1. The levels of each feature, listed in the third column of Table 1, were randomly selected in each task by taking two levels from a randomized list that contained each level twice (e.g., “Not at all,” “Not at all,” “Somewhat,” “Somewhat,” “To a great extent,” “To a great extent”), which made combinations of two different levels slightly more likely and combinations of the same levels slightly less likely than if the feature levels were selected for each artificial being with equal probability. An example choice task is shown in Figure 1. We used the same levels (i.e., “Not at all”, “Somewhat”, “To a great extent”) for many of the features to maintain consistency and limit cognitive load, though they could have been interpreted in different ways for different features.

We asked participants whether they think it could ever be wrong to harm an artificial being that exists either now or in the future (1 = Definitely not, 7 = Definitely). This question was used in sensitivity analysis, reported in the supplementary material. Using the same scale, we also asked participants whether they think artificial beings could ever experience pain or pleasure and whether artificial beings could be as intelligent as a typical human. These latter two questions were collected for exploratory purposes and were not used in any further analysis; we report these results in the supplementary material.

Participants then answered demographic questions on their age, gender, ethnicity, education, income, and political views. These questions were used both to understand the sample characteristics and to test for interaction effects, such as whether the effects of the features on moral consideration differ based on political views with results shown in the supplementary material. Finally, participants were debriefed and given the opportunity to provide feedback on the study.

\section{Results}

\subsection{Individual Feature Effects}

In a conjoint experiment, we are interested in the average marginal component effects (AMCE)—the effects on moral consideration of an AI having a specific feature (e.g., “Somewhat,” “To a great extent”) versus not having that feature \cite{hainmueller14}. These can be estimated with linear regression under testable assumptions \cite{hainmueller14}, which we validate in the supplementary material. Each participant evaluated two descriptions of AIs in 13 choice tasks, so in total 30,238 AIs were evaluated. Since seven of the 11 features were shown per task, we had on average 19,242 data points to estimate the effects of each feature. However, because each participant completed multiple tasks, the data points are not independent. We therefore estimated the effects of the features with standard errors clustered at the participant level. 

The AMCEs are presented in Figure 2 and Table 2. The second column of Table 2 is the estimated effect for each feature. For example, the estimate of 0.062 for “Autonomy: Somewhat” indicates that if an AI was described as being  “somewhat” autonomous, participants were 6.2 percentage points more likely to choose that AI as being more morally wrong to harm than an AI described as “not at all” autonomous. As the table and figure show, each of our 11 hypotheses (H1–H11) were supported; each of the features significantly affected participants choices about which AI it would be more morally wrong to harm in the expected direction. These results remained significant with a correction for multiple comparisons that held the false discovery rate at 10\% \cite{benjamini95}; see Table S5 in the supplementary material.

\begin{figure*}[h!]
    \label{fig:amces}
    \includegraphics[width=0.92\textwidth]{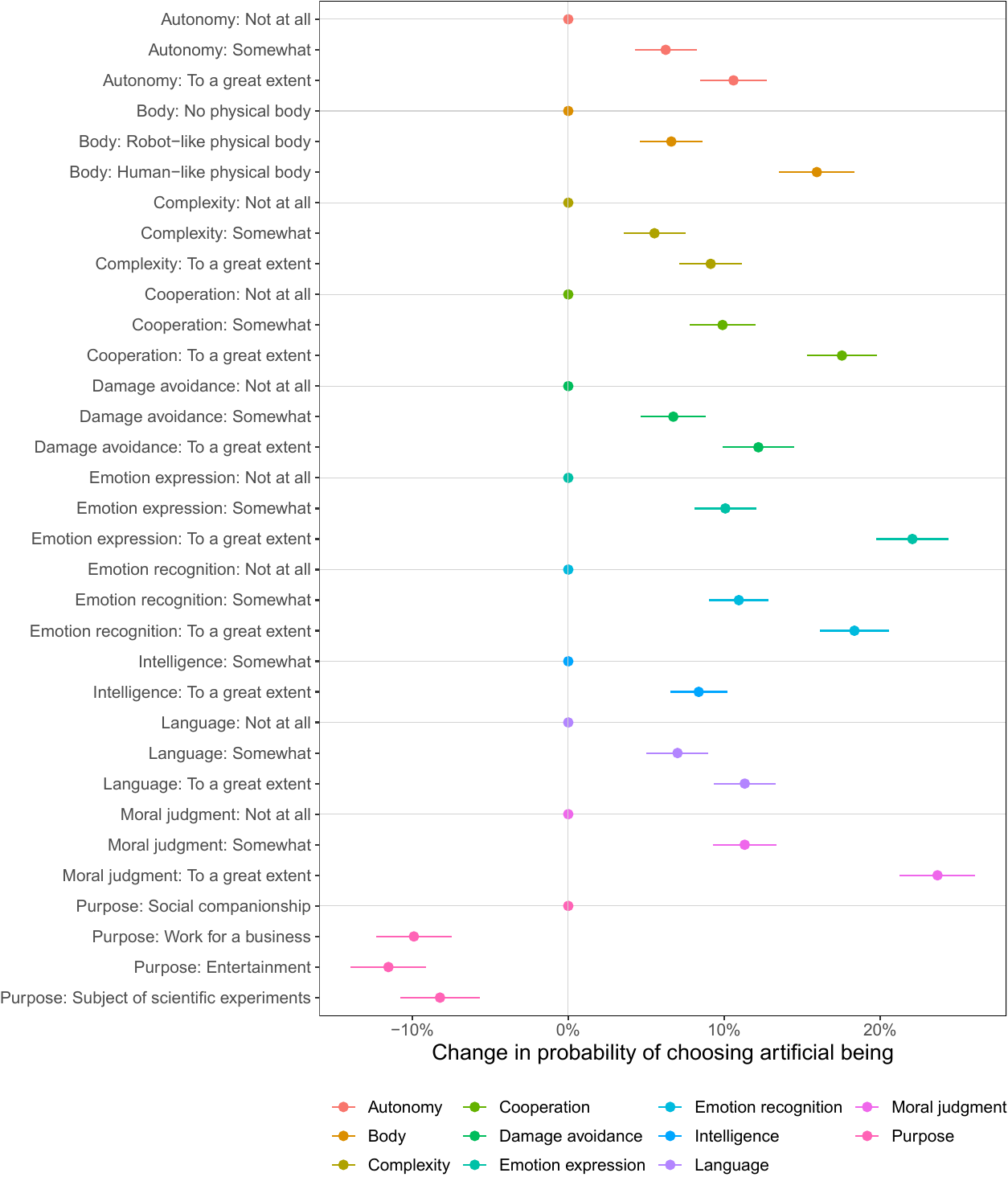}
    \caption{Average Marginal Component Effects. The dots with horizontal bars (color-coded for each feature) represent the means and 95\% confidence intervals of the effects of feature level on the probability of choosing an artificial being as being more wrong to harm relative to the baseline level, which is shown as a dot on the vertical line crossing the x-axis at 0\%. Where the bars do not cross the vertical line at 0\%, the effects can be interpreted as statistically significant. Confidence intervals are calculated based on standard errors clustered at the respondent level.}
\end{figure*}

\subsection{Categories of Effect Sizes}

We conducted pairwise comparisons to test for differences in the size of effects between the features \cite{clogg95, paternoster98}. For the features that were measured on three-point Likert scales (“Not at all,” “Somewhat,” “To a great extent”), we compared the effects of the AI having the feature in question “to a great extent” versus “not at all.” For Body, we compared the effect of the AI having a “human-like physical body” versus “no physical body.” For Purpose, we compared the effect of the AI having a social purpose versus any non-social purpose. We did not include Intelligence in this analysis because, while it was on the same Likert scale as most of the other features, we only included two levels (“Somewhat,” “To a great extent”), as described in the methodology section, making effect size comparisons with the other features particularly difficult. We report the key results here; full results can be found in Table S7 of the supplementary material.

The top two features, Moral Judgment and Emotion Expression, were not significantly different from each other ($b_{diff}$ = 0.02, $Z$ =0.94, $p$ =0.346). The next strongest feature, Emotion Recognition, was significantly less important than both Emotion Expression ($b_{diff}$ = 0.04, $Z$ =2.28, $p$ =0.023) and Moral Judgment ($b_{diff}$ = 0.05, $Z$ =3.19, $p$ =0.001), but was not significantly different from having a human-like physical body ($b_{diff}$ = 0.02, $Z$ =1.44, $p$ =0.149) or Cooperation ($b_{diff}$ = 0.01, $Z$ =0.50, $p$ =0.619). Emotion Recognition, Body, and Cooperation were all significantly more important than all of the remaining features (see the supplementary material for full statistics). There were no significant differences between Damage Avoidance, the next strongest feature, and Language ($b_{diff}$ = 0.01, $Z$ =0.57, $p$ =0.571), Autonomy ($b_{diff}$ = 0.02, $Z$ =1.00, $p$ =0.318), or Purpose ($b_{diff}$ = 0.02, $Z$ =1.45, $p$ =0.145), though Damage Avoidance was significantly more important than the least strong feature, Complexity ($b_{diff}$ = 0.03, $Z$ =1.97, $p$ =0.049). The next strongest feature, Language, was not significantly more important than Complexity ($b_{diff}$ = 0.02, $Z$ =1.51, $p$ =0.132). Some of these differences were no longer significant after multiple comparisons corrections; see the supplementary material for the full statistics. Overall, this analysis suggests that there are broadly three categories of feature effect sizes:

\begin{itemize}
    \item Strongest effects: Moral Judgment, Emotion Expression
    \item Moderately strong effects: Emotion Recognition, Body, Cooperation
    \item Weaker effects: Damage Avoidance, Language, Autonomy, Purpose, and Complexity
\end{itemize}

\section{Discussion}

We conducted a conjoint experiment to estimate the effects of 11 features on the moral consideration of AIs in a single study. As hypothesized, all of the 11 features in our study affected participants’ judgments about the moral wrongness of harming AIs. These results support existing studies that have found positive effects of some of the features included in our study: an AI’s physical body \cite{kuster21, riek09}, emotion expression \cite{lee19, nijssen19}, autonomy \cite{chernyak16, lima20}, damage avoidance \cite{tanibe17, ward13}, intelligence \cite{bartneck07}, moral judgment \cite{flanagan21, swiderska20}, and purpose \cite{wang18}. The present study adds to the literature by providing evidence of the importance of several features that have received less attention: complexity, cooperation, emotion recognition, and capacity for human language.

We compared each pair of effects to each other to estimate their relative strength. We found three categories of effect size. In the first category, with the strongest effects, were an AI’s capacity for moral judgment and emotion expression. In the second category were emotion recognition, cooperation, and having a human-like physical body. In the third category, with the weakest effects, were autonomy, complexity, damage avoidance, language, and having a social purpose. While intelligence also had a positive effect, with the effect of having intelligence “To a great extent” compared to “Somewhat” being of a similar magnitude to the equivalent comparison for the features in the second category (see Table S8 in the supplementary material), we did not formally include it in this analysis because it was measured differently to the other features, as described above. In general, intelligence could be considered a meta-feature that undergirds many of the other features that we considered; it does not seem possible that a being with no intelligence at all could, for example, be autonomous, avoid damage, or recognize emotions in others.

Four of the top five features—emotion expression, emotion recognition, cooperation, and moral judgment—reflect an AI’s capacity to interact prosocially with humans. The extant literature has focused most on the capacity for experience as a driver of moral consideration \cite{gray12}. Why do we instead find prosociality matters most in the case of AIs? This may reflect that humans perceive AIs as threatening—to our resources, our identity, and even our survival \cite{zlotowski17}. We therefore grant them moral consideration conditionally, to the extent that they show prosocial intentions towards us. Further understanding the effects of these prosocial features, especially why they have the strong effects that they do in the context of AI, is a key topic for future research.

Other than prosociality, the strongest effect was having a human-like physical body. This could be explained via an increased perception that the AIs have minds \cite{abubshait17, ferrari16, gray12a}, though this explanation seems less likely because we included a range of features indicative of mind (e.g., emotion expression, damage avoidance) alongside an AI’s body. A second possibility is that it reflects an anthropocentric bias based on mere appearance and human-likeness, perhaps echoing work in HRI \cite{hou23}, human-agent interaction \cite{chavezgonzalez22}, and social psychology \cite{ladak23} that shows humans also engage in group-based dynamics, such as in-group favoritism, with AIs. These possible explanations should be tested in future research.

From a design perspective, we know that AIs with human-like physical bodies and prosociality can promote better quality HCI \cite{eyssel10, zlotowski15}. This can be due to factors such as creating greater familiarity with the AI and building on existing skills developed in social interactions between humans \cite{zlotowski15}. The present study suggests that building AIs with human-like bodies and prosociality may have significant effects on moral consideration. Given the importance of morality in social interaction, designers may want to implement such features in AIs only when they aim to mimic human-human interaction. By increasing moral consideration, designing AIs with human-like bodies and prosociality could also help solve the problem of people being abusive towards AIs \cite{angeli08, nomura15}, which can cause expensive damage and dangerous situations for bystanders, though further research should be conducted on this question because human-likeness in AIs has also been found to be associated with greater levels of abuse \cite{keijsers21}. Additionally, Schwitzgebel and Garza \cite{schwitzgebel15} argue that we should design AI systems that evoke reactions that reflect their true moral status (i.e., how much they matter morally, for their own sake). If we build AIs with capacities associated with moral status, such as consciousness \cite{ladak23c} or sentience \cite{anthis21}, we should consider also designing them with human-like bodies, prosociality, or other features that affect moral consideration to facilitate accurate perceptions of the AIs. On the other hand, they argue that if the AIs do not actually have moral status, then building them with consideration-provoking features could result in people wasting resources to benefit AIs that they erroneously think warrant moral consideration. Another consideration against evoking such reactions is that they can cause psychological distress and conflict in users who feel that they have obligations towards the AIs \cite{laestadius22}. Overall, AI designers should consider that building AIs with certain features will likely have effects on moral consideration with a variety of consequences for interaction, sometimes unintended.

\section{Limitations}
\balance

Our study has some limitations. First, while the Prolific sample had some demographic measures close to the U.S. population (e.g., 47.9\% women), it was not nationally representative, and we did not collect data from outside the U.S.

Second, conjoint experiments test hypothetical preferences rather than real-world behaviors. While such information is important, and many societal decisions are made on the basis of such hypotheticals (e.g., voting for social policies), they do not always translate to practical behavior, such as in the privacy paradox, the finding that people consistently report preferences for privacy that are not borne out in their online behavior \cite{kokolakis17}. Future research should test the relative effects of these features in more concrete scenarios, such as with large language models, interactive robots, virtual agents, and other multifunctional AI systems. 

Third, we asked participants how morally wrong they considered it to harm AIs. While this is a core aspect of moral consideration \cite{gray12a}, moral consideration arguably has additional aspects, such as the attribution of rights. Also, while we gave participants background information about this idea, the use of a single measure is more likely to be misinterpreted than a more detailed measure would be. For example, participants could have interpreted our question in terms of the wrongness of actions they could take against the AIs (e.g., kicking a physical robot vs. deleting a non-physical AI) rather than about the AIs themselves. To explore this further, we conducted a study with 20 new participants asking why they thought it was morally wrong to harm the AIs they chose in this task and what they understood by the word “harm.” As detailed in the supplementary material, participants tended to give reasons relating to the AIs themselves rather than specific actions (e.g., almost 50\% indicated choosing AIs that had features that made them seem more human). Participants also typically understood the word “harm” broadly, capturing any sort of damage to the AIs, physical or psychological (e.g., “to injure, inflict pain, inflict physical or mental violence.”) Overall, it seems that participants interpreted the question as we intended. Still, future research should assess additional aspects of moral consideration, such as through Piazza et al.’s moral standing scale \cite{piazza14}.

Fourth, we used the levels “Not at all,” “Somewhat,” and “To a great extent” to describe the way in which the AIs had most of the features. While these levels are intended to be neutrally worded, it may be that, for example, people perceive the word “somewhat” differently when paired with “complex” compared with “intelligent.” This is important to be aware of when making comparisons across features. An alternative approach would be to use feasture levels that are tailored to the specifics of each feature, though this could increase cognitive load, and, at least in the present study, it would introduce additional variation that makes direct comparisons more challenging. Future research should test such alternative designs.

Finally, our study prioritizes breadth over depth. This means that our operationalizations have less nuance than they would in a study of only a small number of features. For example, we operationalized “autonomy” as varying along a single dimension, the degree of independence from human control, but autonomy is more complicated, such as in the type of human control exerted. Similarly, we operationalized “body” using only three levels, “Human-like physical body,” “Robot-like physical body,” and “No physical body,” but there are other possibilities, such as a zoomorphic body or an ability to be uploaded into different bodies. There are many openings for future studies to build on this breadth-focused study by exploring particular variations across and within these features, especially of the features with the largest measured effects reported here.

\section{Conclusion}

AI systems are increasingly evoking moral reactions from humans. Because AIs can have a wide range of relevant features, we conducted an experiment testing the effects of 11 features on the moral consideration of AI. The presence of each of the features increased moral consideration, with the strongest effects from having a human-like physical body and the capacity for prosociality. In a world where AIs are perceived as threatening to humans, such as by replacing us in the workplace and challenging our sense of uniqueness, the highest levels of moral consideration may only be granted if the AI shows positive intentions.

\begin{acks}
We would like to thank Kirk Bansak, Janet Pauketat, and Yanyan Sheng for their thoughtful feedback on various aspects of this project.
\end{acks}

\bibliographystyle{ACM-Reference-Format}
\bibliography{bibliography/which_ais}

\end{document}